\documentclass[USenglish,twocolumn]{article}	

\usepackage[utf8]{inputenc}				
\usepackage[big,online]{dgruyter}	
\usepackage{lmodern} 
\usepackage{microtype}

\usepackage[sorting=none,firstinits=true,style=numeric-comp,
url=false,eprint=false,isbn=false,maxbibnames=6,minbibnames=3]{biblatex}
\renewbibmacro{in:}{}
\DeclareFieldFormat{pages}{#1}
\usepackage{hyperref}

\begin{document}

	\articletype{Article}

  \startpage{1}
  \aop

\title{Reaction Nanoscopy of Ion Emission from Sub-wavelength Propanediol Droplets}
\runningtitle{Droplet Reaction Nanoscopy}

\author[1,2,7]{Philipp Rosenberger}
\author[1,2]{Ritika Dagar}
\author[1,2,3]{Wenbin Zhang} 
\author[4]{Arijit Majumdar}
\author[1,2]{Marcel Neuhaus}
\author[4,6]{Matthias Ihme}
\author[1,2]{Boris Bergues}
\author[1,2,5,6,7]{Matthias F. Kling}
\runningauthor{P.~Rosenberger et al.}
\affil[1]{\protect\raggedright 
Department of Physics, Ludwig-Maximilians-Universit\"at Munich, D-85748 Garching, Germany}
\affil[2]{\protect\raggedright 
Max Planck Institute of Quantum Optics, D-85748 Garching, Germany}
\affil[3]{\protect\raggedright 
State Key Laboratory of Precision Spectroscopy, East China Normal University, Shanghai 200241, China}
\affil[4]{\protect\raggedright 
Department of Mechanical Engineering, Stanford University, Stanford, CA 94305, USA}
\affil[5]{\protect\raggedright 
Department of Applied Physics, Stanford University, Stanford, CA 94305, USA}
\affil[6]{\protect\raggedright 
SLAC National Accelerator Laboratory, Menlo Park, CA 94025, USA}
\affil[7]{\protect\raggedright 
Corresponding authors, e-mails: philipp.rosenberger@physik.uni-muenchen.de, kling@stanford.edu}
	
\abstract{Droplets provide unique opportunities for the investigation of laser-induced surface chemistry. Chemical reactions on the surface of charged droplets are ubiquitous in nature and can provide critical insight into more efficient processes for industrial chemical production. Here, we demonstrate the application of the reaction nanoscopy technique to strong-field ionized nanodroplets of propanediol (PDO). The technique’s sensitivity to the near-field around the droplet allows for the in-situ characterization of the average droplet size and charge. The use of ultrashort laser pulses enables control of the amount of surface charge by the laser intensity. Moreover, we demonstrate the surface chemical sensitivity of reaction nanoscopy by comparing droplets of the isomers 1,2-PDO and 1,3-PDO in their ion emission and fragmentation channels. Referencing the ion yields to gas-phase data, we find an enhanced production of methyl cations from droplets of the 1,2-PDO isomer. Density functional theory simulations support that this enhancement is due to the alignment of 1,2-PDO molecules on the surface. The results pave the way towards spatio-temporal observations of charge dynamics and surface reactions on droplets in pump-probe studies.}


\maketitle

\section{Introduction} 

Over recent years, it has been repeatedly demonstrated that chemical reactions on charged droplets can be orders of magnitude more efficient compared to bulk reactions\,\cite{lee2015,Yan2016,Lai2018,Ruiz-Lopez2020,Burris2021}. 
This makes charged droplets a possible candidate for the origin of prebiotic polymers\,\cite{Griffith2012,Dobson2000,Gale2020} and relevant for the chemical and pharmaceutical industry\,\cite{Nie2020,Zhong2020}. 
The underlying mechanisms responsible for the accelerated reactions are, however, still debated.
While earlier studies suggested solvent evaporation as the main explanation for the enhancement\,\cite{BaduTawiah2012,Bain2015,Girod2011}, recent studies emphasize the role of the droplet surface\,\cite{lee2015,Marsh2019,Li2016}.
The surface gives rise to high molecular concentrations and a high degree of molecular alignment\,\cite{Xiong2020,Jung2007,Griffith2012,Dobson2000,Gale2020}, a high charge density and acidity\,\cite{Sahraeian2019,Banerjee2015,Huang2021}, as well as a strong electric field\,\cite{Xiong2_2020,Kwan2020,Hao2022}.
While all of these effects are important factors in the acceleration of chemical reactions, elucidating their individual contributions is still a subject of active research\,\cite{Ruiz-Lopez2020,Hao2022}.

In experimental studies, charged droplets are usually generated from electrospray ionization sources.
Despite the widespread use of such sources, the ion emission mechanisms from electrospray-generated droplets are still not fully understood\,\cite{Conner2021,Konermann2013}.
A major reason is the complexity of the electrospray ionization mechanism itself. 
Electrospray droplets have wide distributions in both size and charge that strongly depend on the electrospray source\,\cite{Conner2021,Luebbert2021}. 
In combination with solvent evaporation, droplets generated by electrospray may become unstable as their size becomes smaller and their charge approaches the Rayleigh limit\,\cite{Rayleigh1882}:
\begin{equation}
\label{eq:Rayleigh}
    q_R = \pi\sqrt{8\gamma\varepsilon_0 d^3},
\end{equation}
where $\gamma$ is the surface tension, $d$ is the droplet diameter, and $\varepsilon_0$ is the vacuum permittivity. 
Due to this complexity, a simpler injection system with a more well-defined initial state is beneficial for the study of droplet chemistry.

\begin{figure*}
	\centering\includegraphics[width=0.68\textwidth]{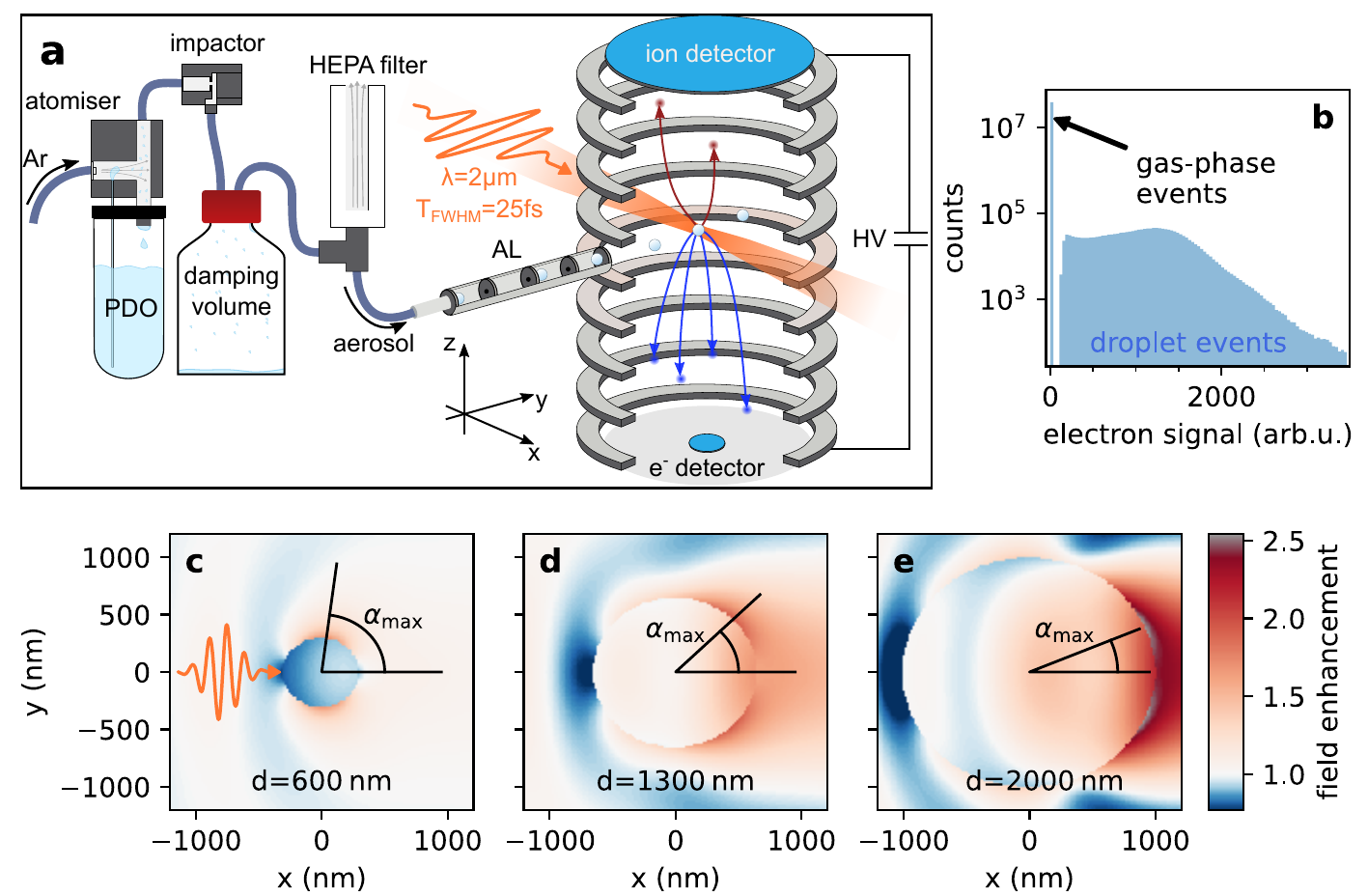}
	\caption{\label{Fig:FigureSetup} 
 {\bf a:}~Droplet source and reaction nanoscope. The propanediol (PDO) droplet source consists of an atomizer operated with argon (Ar), an impactor, a damping volume (1\,l), and a high-efficiency particulate air (HEPA) filter. An aerodynamic lens (AL) was used to inject a collimated aerosol stream into the reaction nanoscope where the droplets were ionized in the focus of the laser (polarization along $y$). The electrons and ions were directed to their detectors by an electric field generated by a high voltage (HV).
 {\bf b:}~Electron signal distribution. The signal is the time integral of the voltage pulse from the $e^-$ detector.
 {\bf c:}~The pulse symbol indicates the propagation direction ($x$) and the polarization direction ($y$) and of the laser.
 {\bf c-e:}~Field enhancement factors calculated from the Mie solution\,\cite{Mie1908} as a function of position in the $x$-$y$ plane for 1,2-PDO droplets at the origin with different diameters $d$. The angle $\alpha_{\rm max}$ is the angle of maximum field enhancement at a distance of 1\,Å from the droplet surface.
 }
\end{figure*}

This is addressed by the application of reaction nanoscopy to droplets. The technique holds great potential for the control and direct observation of nanodroplet surface reactions and may help to elucidate the mechanisms behind the enhancement of reaction rates.
Reaction nanoscopy enables the measurement of the three-dimensional momentum distribution of ions generated by a strong laser pulse from the surface of an isolated nanosystem injected into ultra-high vacuum (see \autoref{Fig:FigureSetup}a). 
For solid silica nanoparticles, it was demonstrated that the laser-induced proton emission strongly correlates with the optical near-field distribution on the particle surface\,\cite{Rupp2019,Rosenberger2020}, and that the emission region on the surface can be controlled on a nanometer scale by tailored laser fields\,\cite{Zhang2022}.
The technique was recently shown to provide information about surface chemistry in a study on the formation of trihydrogen cations from water-covered isolated silica nanoparticles\,\cite{Alghabra2021}.

Extending these prior investigations, the objective of this work is to demonstrate the application of reaction nanoscopy to the liquid phase using polydisperse droplets of PDO as the model system. We analyze the ion spectra from 1,2-PDO and 1,3-PDO targets, and demonstrate that reaction nanoscopy's near-field sensitivity is preserved for droplets allowing in-situ characterization of the droplets size and charge.
The droplet-specific reaction nanoscopy data indicate a high relevance of fragment protonation for the generation and emission of ions.
Furthermore, by comparing the single-molecule fragmentation of the PDO isomers to the ions emitted from the PDO droplets,
we find that the molecular alignment on the surface of 1,2-PDO droplets enhances the production of  CH$_3^+$ ions.
This interpretation is confirmed by density functional theory (DFT) calculations.

\section{Experimental}

\subsection{Laser system}
We used a home-built laser system based on optical parametric chirped-pulse amplification\,\cite{Neuhaus2018}.
The system features a central wavelength of 2\,$\mu$m, a maximum pulse energy of 100\,$\mu$J, a pulse duration of 25\,fs and a repetition rate of 100\,kHz. 
The pulse energy was adjusted by a combination of a broadband half-wave plate and a wire-grid polarizer. 
The pulses were compressed by maximizing the ion count rate detected in the reaction nanoscope while changing the dispersion with a pair of fused silica wedges.
Inside the nanoscope, the pulses were back-focused using a spherical silver mirror with a focal length of 75\,mm. 
The peak intensity in the interaction region reached up to 5$\times$10$^{13}$\,W/cm$^2$.

\subsection{Droplet source}
In order to investigate the influence of the molecular structure on the ion emission from the droplet surface, we compared droplets of  1,2-PDO (Thermo Scientific, 99.5\,\%) and 1,3-PDO (Thermo Scientific, 99\,\%). Since both liquids are highly hygroscopic, all experiments were carried out with samples from freshly opened bottles.
The droplet source is shown in \autoref{Fig:FigureSetup}a.
We produced a polydisperse PDO-argon aerosol using an atomizer (TSI inc., model 3076) with 20\,psi (1.4\,bar) of argon pressure for 1,3-PDO. 
In order to aerosolize 1,2-PDO, a pressure of 30\,psi (2.1\,bar) was required. 
The polydispersity was reduced by an impactor (TSI inc., part no. 1035900) with an aperture of 0.71\,mm (TSI inc., part no. 390170) and a gap of 2\,mm. As instructed by the atomizer manual, a 1\,l bottle was used after the impactor to dampen flow instabilities and to collect excess PDO from the impactor.
After the damping volume, a high-efficiency particulate air (HEPA) filter connected to the aerosol line on the inlet side and to ambient air on the outlet side was used to ensure a pressure of 1\,atm inside the line.
The pressure of 1\,atm is required at the inlet of the aerodynamic lens, which was used to introduce the aerosol stream into the ultra-high vacuum of the reaction nanoscope.
The aerodynamic lens collimated the droplet stream and further reduced the polydispersity\,\cite{Bresch2007}.
The lens assembly was followed by three stages of differential pumping to minimize the gas load in the experimental chamber. 
The average size of the nanodroplets was determined to be (590$\pm$50)\,nm (see results section).

\subsection{Reaction nanoscopy with droplets}
The reaction nanoscopy technique was described in previous publications\,\cite{Rupp2019,Rosenberger2020,Alghabra2021,Zhang2022,Rosenberger2022}.
Briefly, a strong laser pulse ionizes a nanoscale target and the released ions and electrons are forced towards detectors on opposing ends of a spectrometer by a static homogeneous electric field, as shown in \autoref{Fig:FigureSetup}a.
The ion detection and momentum reconstruction schemes are the same as for the reaction \emph{microscopy} technique\,\cite{Ullrich2003}, where the ion recoil momentum is reconstructed from the time-of-flight and the arrival position of the ions, which are detected by a combination of a microchannel plate stack with a delay-line anode.

\begin{figure*}[htb!]
	\centering\includegraphics[width=0.7\textwidth]{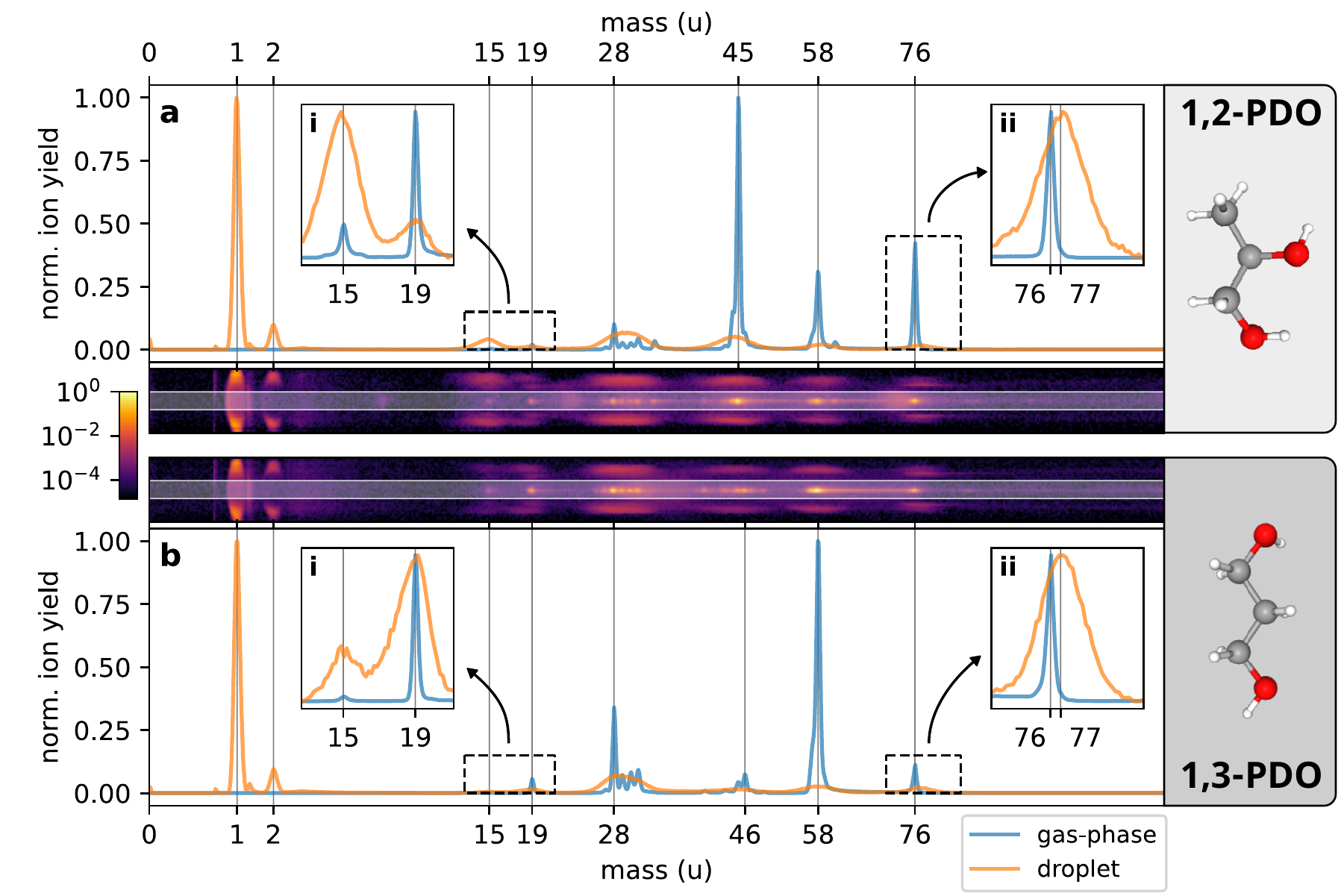}
	\caption{\label{Fig:1D_tof} 
   {\bf a:}~Time-of-flight spectra (in mass units) for ions generated from single gas-phase molecules (blue) and on the surface of droplets (orange) obtained from a single measurement with 1,2-PDO at a laser intensity of approx. 5$\times$10$^{13}$\,W/cm$^2$. The insets {\bf i} and {\bf ii} contain a zoomed-in and normalized view of the mass ranges indicated by the dashed boxes. The mass of the PDO molecule (or PDO$^+$ ion) is $m=76\,$u.
    {\bf b:}~Same as panel a using 1,3-PDO as the target.
    Gas-phase and droplet data were separated based on the electron signal (see \autoref{Fig:FigureSetup}b). The droplet data was further cleaned from gas-phase ions by removing ions with a small deflection from the center of the ion detector along the polarization direction of the laser. 
    {\bf pseudo-color~plots:}~Position-resolved (along polarization direction $y$) time-of-flight histograms for ions with high electron signal. Only the counts outside the white-shaded regions ($|y|>11$\,mm) are contained in the droplet histograms of panel a and b. The time-of-flight histogram for counts with $|y|\leq11$\,mm is practically identical to the gas-phase case (see supplementary material).
 }
\end{figure*}

The electron detector of the reaction nanoscope is a channeltron electron multiplier. The voltage signal from the channeltron is time-integrated for every laser shot and used as a measure of the number of emitted electrons. 
The integrated electron signal is used to assign all ions of a laser shot to either the ionization of a single gas-phase molecule or to the ionization of a droplet (see \autoref{Fig:FigureSetup}b). 
A small or zero value is indicative of a gas-phase ionization event, whereas a large value corresponds to the ionization of a droplet, which releases many electrons. The ideal threshold value is determined by a gas-only measurement in the absence of droplets. 
This categorization of laser shots helps to remove pure gas-phase ionization events from the droplet data. However, due to the partial evaporation of the droplets in vacuum and the large focal volume compared to the droplet size, the ionization of a droplet is frequently accompanied by ions from single molecules. These are identified by a comparison with the gas-phase data, and are rejected based on their arrival position on the detector as shown in \autoref{Fig:1D_tof} and the supplementary material.

\section{Theoretical}
To interpret the experimental results, we performed DFT calculations for the minimum energy configuration for bulk 1,2-PDO and 1,3-PDO. The simulations were performed with VASP\,\cite{Kresse1993,Hafner2008}, which uses the plane wave DFT formulation\,\cite{Jensen2017}. 
The exchange correlation term is handled using the Perdew-Burke-Ernzerhof functional\,\cite{Perdew1996}, and the Projector Augmented Wave method\,\cite{Bloechl1994} is used to represent the interaction between atom cores and electrons.
To describe the interface at the droplet surface, we consider a planar slab of molecules, neglecting curvature effects on the equilibrium interface structure owing to the large droplet radius. 
The simulation cell is 8Å$\times$90Å$\times$12Å in dimension and has periodic boundaries along the direction of the surface interface. 
Along the surface-normal direction, we approximate the liquid phase by six layers of molecules with rotational symmetry around the center of the simulation cell. See the supplementary material for more details. The symmetry constraint is used for an unambiguous definition of the droplet-vacuum interface.
The edge length of the cell in the normal direction is kept sufficiently large to avoid any effect of periodicity. A 7$\times$1$\times$7~$\Gamma$-centered mesh is used to sample the Brillouin zone. For energy minimization, the energy convergence criterion is set to ${10}^{-6}$\,eV and the norm of the forces acting on the ions is less than 0.01\,eV/Å.

\begin{figure*}
    \centering
    \includegraphics[width=\textwidth]{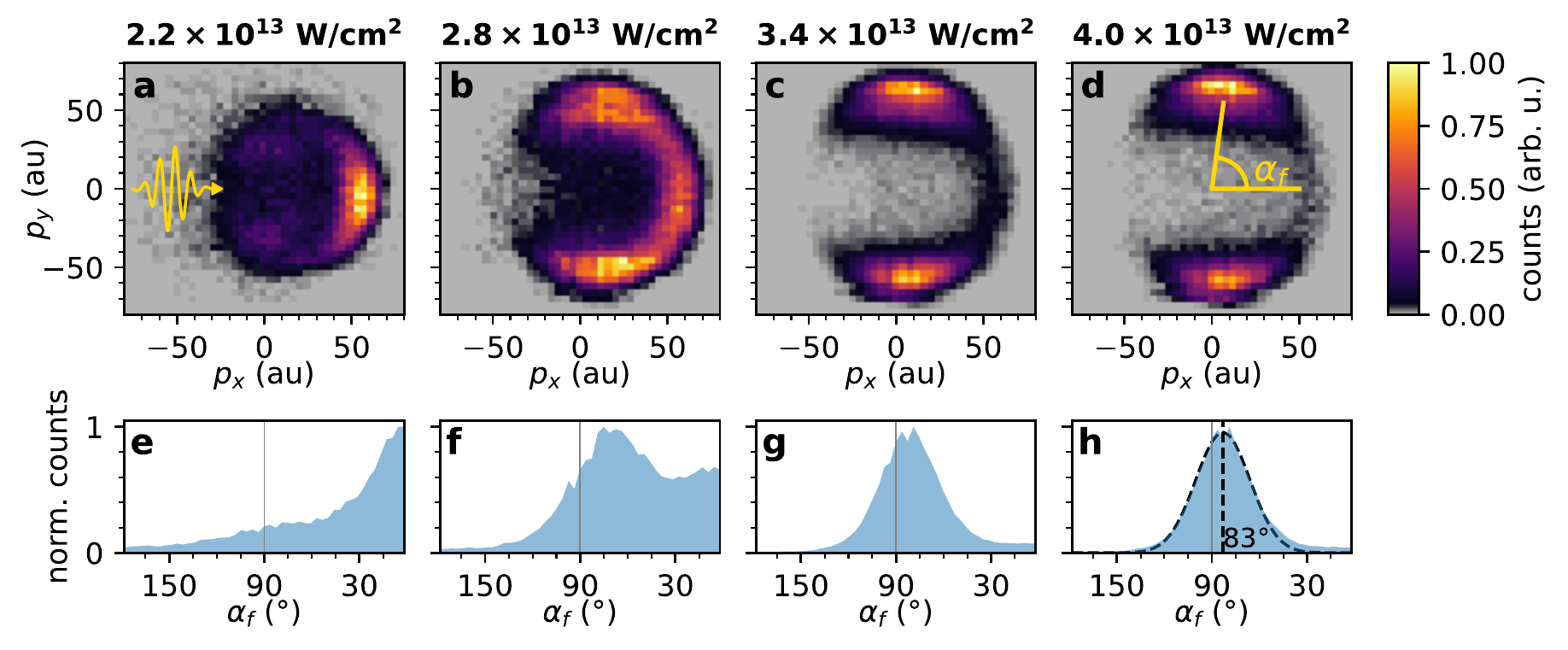}
    \caption{ \label{fig:intensityScan}
Intensity-dependent proton emission from polydisperse droplets. 
    {\bf a:}~The pulse symbol indicates the propagation direction ($x$) and the polarization direction ($y$) and of the laser.
    {\bf a-d:}~Proton momentum histograms for the plane $p_x$-$p_y$ for different laser intensities. The momenta are given in atomic units (au).
    {\bf e-h:}~Distributions of the final proton emission angle $\alpha_f$ with respect to the laser propagation axis in the $x$-$y$ plane, defined as $\alpha_f=\arctan(|p_y|/p_x)$, as indicated in panel d.
    {\bf h:}~The dashed vertical line shows the mean (83$^\circ$) of a Gaussian fit (dashed curve) to the angular distribution, representing the most likely proton emission angle.
    }
\end{figure*}

\section{Results and Discussion}
\subsection{Intensity-dependent near-field response}
The time-of-flight spectra for typical measurements of 1,2-PDO and 1,3-PDO are shown in \autoref{Fig:1D_tof}. While we observe a multitude of fragments, we will first concentrate our analysis on the most abundant species in the droplet spectra, the protons, and discuss the other fragments in detail in the later sections.
The high abundance of protons is analogous to previous reaction nanoscopy studies on solid, monodisperse nanoparticles\,\cite{Rupp2019,Rosenberger2020,Alghabra2021,Rosenberger2022}. These studies demonstrated that the local near-field on the particle surface and the angular distribution of the proton yield were highly correlated and that the birth angles on the particle surface are practically identical with the final proton detection angles\,\cite{Zhang2022}.\footnote{The birth and detection angles have to be defined in the same coordinate system with the center of the particle at the origin.}
The same effect would be expected for a \emph{monodisperse} stream of nanodroplets. In our experimental setup, however, the droplet size distribution is not easily controllable. It is determined by the atomizer (see, for example, Ref.~\,\cite{Parmentier2021}), the size selective elements further downstream (the impactor and the aerodynamic lens) and the evaporation of molecules from the droplet surface due to the ultra-high vacuum of the instrument.
Since the optical response of a sub-wavelength nanosphere (particle or droplet) strongly depends on its size (\autoref{Fig:FigureSetup}c-e), and the proton emission is a highly non-linear process, the proton angular distribution in reaction nanoscopy follows the near-field distribution\,\cite{Rupp2019}. 
In order to investigate how this affects the averaged proton momentum spectra for the polydisperse droplet stream, we carried out a series of measurements on 1,2-PDO at different laser intensities. Panels a-d of \autoref{fig:intensityScan} contain the corresponding momentum distribution of protons emitted from PDO droplets.
At low intensity, we find that the protons are mostly emitted in the forward direction along the laser propagation direction ($p_x>0$), and that there is only a minor contribution of protons emitted along the polarization direction of the laser ($p_y$) which is distributed in two lobes and symmetric with respect to $p_y=0$. This dipolar part of the emission pattern becomes more dominant as the intensity is increased, shifts to higher momenta (\autoref{fig:intensityScan}b,c) and eventually stabilizes (\autoref{fig:intensityScan}d). Simultaneously, the relative strength of the momentum component in the forward direction diminishes.
We interpret this observation in terms of the difference between the maximum field enhancement factors of droplets of different sizes. As shown for a set of exemplary sizes in \autoref{Fig:FigureSetup}c-e, the field enhancement factor increases with increasing droplet size. 
In combination with the highly nonlinear process of strong-field ionization and proton emission\,\cite{Rupp2019,Rosenberger2022}, this causes a size-selectivity as a function of laser intensity where ionization and ion emission only take place on the larger droplets in the size distribution when the intensity is low. That larger droplets (with diameters on the order of the laser wavelength), however, only make up a small fraction of the aerosol stream is seen in the high-intensity measurement (\autoref{fig:intensityScan}d) where their contribution is negligible.

\subsection{Droplet size characterization}
\begin{figure}[htbp!]
	\centering\includegraphics[width=\columnwidth]{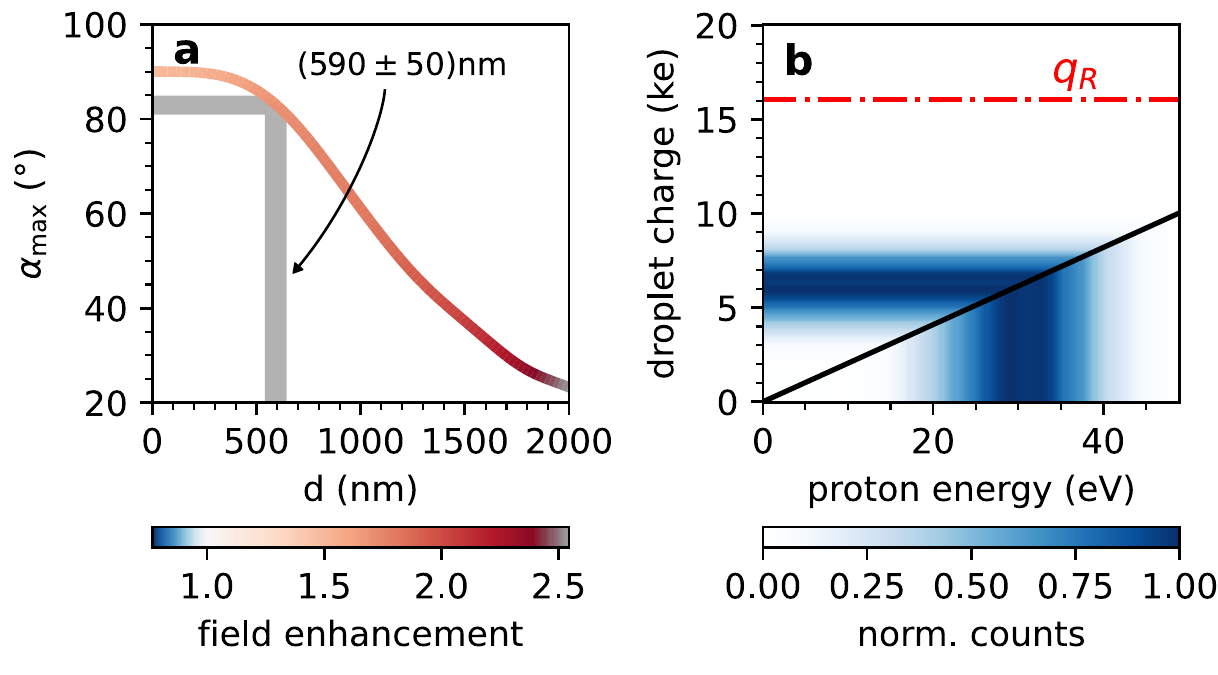}
	\caption{\label{Fig:FigureSize} 
Droplet characterization. 
 {\bf a:}~Relation between the droplet diameter $d$ and $\alpha_{\rm max}$ calculated from the Mie solution by numerical optimization. The color of the line indicates the max. field enhancement for every droplet diameter. The shaded region in gray indicates the estimate of the most frequent droplet size in our experiment ($(590\pm50)\,$nm) based on the angular distribution of proton momenta (\autoref{fig:intensityScan}h). See main text for details.  
 {\bf b:}~The black solid line shows the relation between ion energy and droplet charge for a homogeneous charge distribution. The color map shows the measured distribution of proton energies (below the line) and the corresponding estimate for the droplet charges (above the line). The red dashed-dotted line indicates the Rayleigh charge limit $q_R$ for a 1,2-PDO droplet with a diameter of 590\,nm.
 }
\end{figure}

The fact that the main peak of the proton momentum distribution stabilizes for laser intensities above approx. 3$\times$10$^{13}$\,W$/$cm$^2$ suggests that, beyond this point, a significant majority of droplet sizes in the aerosol stream are ionized and emitting ions. As shown in the bottom panels of \autoref{fig:intensityScan}, the peak of the angular distribution in the $p_x$-$p_y$ plane stays at a constant value of $\alpha_f=83^\circ$ in this regime. Based on the one-to-one mapping between an ion's birth angle and its final angle\,\cite{Zhang2022}, we can perform an angle-based estimate of the most prominent droplet size in the aerosol stream. On a droplet, the most likely birth angle $\alpha_{\rm max}$ of an ion is simply determined by the point of maximum field enhancement on the surface(see \autoref{Fig:FigureSetup}c-e). 
For a certain range of droplet diameters, the angle $\alpha_{\rm max}$ can be related to the droplet size, as shown in \autoref{Fig:FigureSize}a.
The relation between the angle $\alpha_{\rm max}$ and the droplet diameter is obtained from the Mie-Solution of Maxwell's equations (see, for instance, Refs.~\,\cite{Mie1908,BohrenHuffman}) for PDO spheres of different sizes. 
Carrying out this size estimation with the experimental data taken at a laser intensity of approx. 4.0$\times$10$^{13}$\,W$/$cm$^2$, we obtain a mean droplet diameter of $\langle d \rangle = (590\pm50$)\,nm. The error of $\Delta d =\pm50\,$nm was determined based on an uncertainty of $\Delta \alpha_f=\pm2^\circ$ for the determination of the peak of the proton angular distribution (see \autoref{fig:intensityScan}h).

\begin{figure*}[htb!]
	\centering\includegraphics[width=0.7\textwidth]{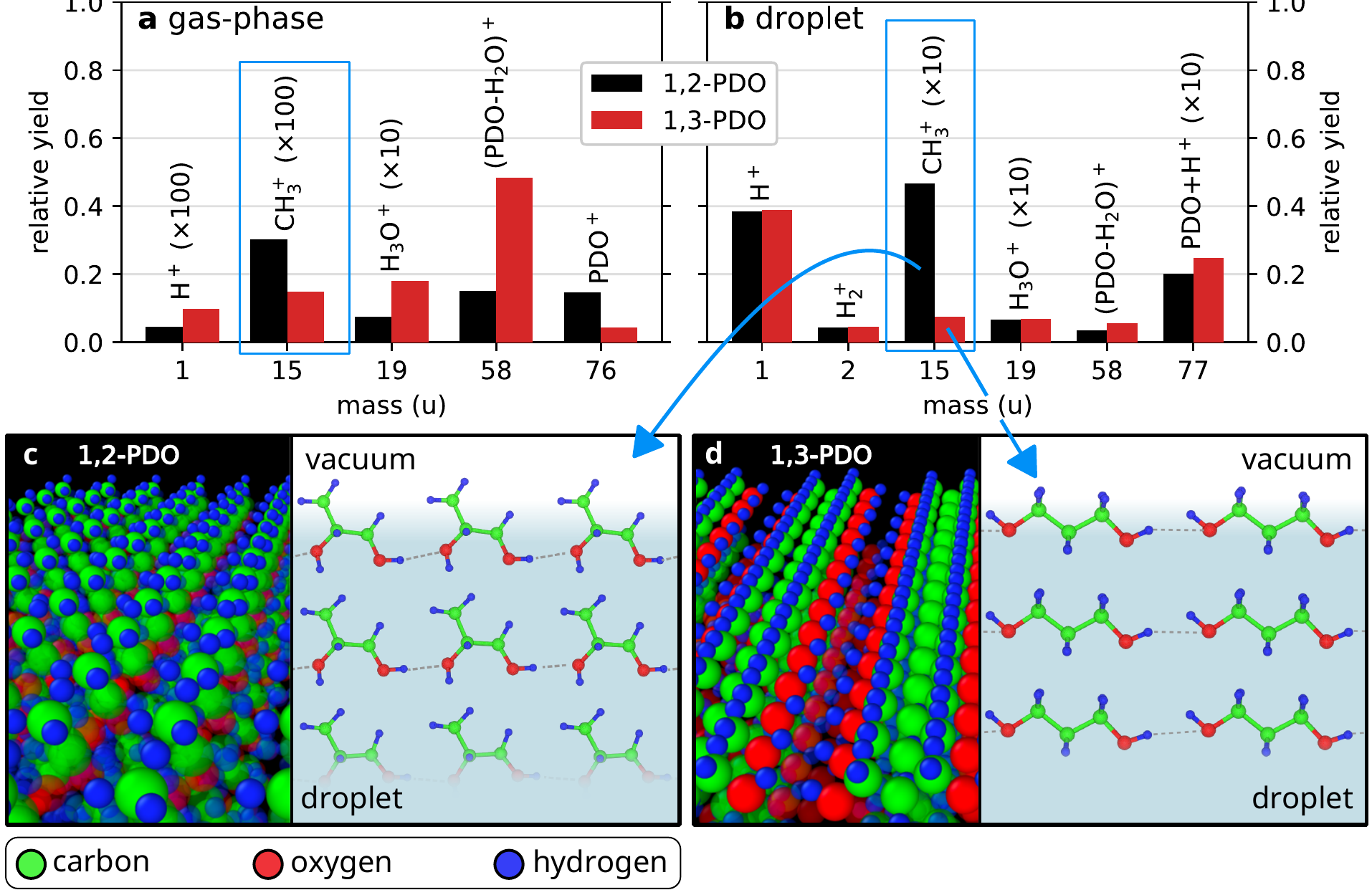}
	\caption{\label{Fig:FigureIsomer} Isomer comparison of the fragmentation channels for gas-phase PDO and PDO droplets. {\bf a:}~The ion yield from gas-phase 1,2-PDO and 1,3-PDO for selected fragments. The data was obtained by filtering the droplet measurements on events with a low electron signal. The yield is normalized to the total number of counts after filtering. {\bf b:}~The same measurements as in panel~\textbf{a} filtered on droplet events (high electron signal). The data is normalized to the total number droplet ions, i.e. ions with a high electron signal. The raw data for panels~\textbf{a} and \textbf{b} can be found in the supplementary material. The blue rectangles mark the CH$_3^+$ ion for which we find a significant difference between the gas-phase and droplet data. We attribute this to the molecular alignment of the PDO molecules on the droplet surface (see text for details). Panels~\textbf{c} and \textbf{d} present density functional theory results for the molecular alignment of both isomers. Only the top three layers are shown. The gray dotted lines represent hydrogen bonds.}
\end{figure*}

\subsection{Droplet charge and stability}
The results presented here demonstrate that the concepts of reaction nanoscopy with solid nanoparticles can also be applied to droplets. This suggests that in the range of intensities investigated here, the physics of the ion emission process from droplets are comparable to the case of solid nanospheres. Solid nanoparticles are, however, very stable and can hold a high (surface) charge without disintegrating. In contrast, for droplets this is not necessarily the case and may require different modeling.

The stability of charged droplets can be determined from the Rayleigh charge limit (see Eq.~\eqref{eq:Rayleigh}).
Its derivation is based on the idea that the maximum amount of charge a droplet can hold is limited by its surface tension. Once the Coulomb energy $E_C=q^2/8\pi\varepsilon_0R$ of a droplet with radius $R$ exceeds the surface energy $E_S=4\pi\gamma R^2$ (with surface tension $\gamma$) by a factor of two ($E_C>2E_S$), the droplet is no longer stable and will start fragmenting into smaller droplets\,\cite{Kronermann2009,Naher1997}.
Even though the Rayleigh limit does not provide an exact criterion as to when droplet fission sets in\,\cite{Duft2003}, droplet charges far below and above the limit can be safely called stable or unstable, respectively.
Evaluating the stability of the droplets in our experiment requires knowledge about their total surface charge. Since the final ion energy measured with the reaction nanoscope is dominated by the Coulomb repulsion of the ions from the charge on the surface\,\cite{Rupp2019,Rosenberger2020}, a charge estimate can be directly obtained from:
\begin{equation}
    q = E4\pi\varepsilon_0 R/e.
\end{equation}
Here, $E$ is the ion energy obtained from measured ion momentum. We neglected the inhomogeneous charge distribution due to the near-field and assumed a homogeneously charged droplet instead. The assumption of a homogeneous charge density overestimates the total droplet charge, which is acceptable for determining droplet stability as long as the estimated charge turns out to be below the Rayleigh limit.

In \autoref{Fig:FigureSize}b, we show a charge estimate using the energy distribution of protons emitted from 1,2-PDO droplets. 
We find that the distribution of droplet charges is centered at approximately 6000\,e and does not overlap with the Rayleigh charge limit for a 590\,nm PDO droplet of $q_R\approx$16000\,e. 
We therefore conclude that the droplets studied here did not undergo droplet fission after the ionization by the laser pulse.

We observed slightly different energy distributions for different ion species and therefore used the most energetic ions, the protons, as an upper limit for determining the droplet stability.

\subsection{Ion emission: gas-phase vs. droplet}

The use of droplets for reaction nanoscopy comes at the advantage that the material of bulk and surface are identical which makes the origin of the emitted ions unambiguous. For solid particles, this is in general not the case\,\cite{Rosenberger2022}. Besides, the reaction nanoscope can be used for studying single molecules in the gas-phase allowing for a direct comparison to the ion generation and emission from the droplet.
We next discuss this comparison and then focus on isomer-related effects.
We restrict the discussion to unambiguously identifiable ion species. The design of the reaction nanoscope as a momentum imaging spectrometer results in wide time-of-flight distributions for energetic ions, which leads to overlapping times-of-flight spectra for ions with similar $m/q$-values. This is illustrated by the fragments at masses close to $m=28$\,u (C$_2$H$_n^+$ and CH$_n$O$^+$) and $m=45$\,u  (C$_3$H$_n^+$ and C$_2$H$_n$O$^+$)  in \autoref{Fig:1D_tof} where the exact number of H atoms cannot be determined. We excluded such ambiguous species from our analysis.

\autoref{Fig:FigureIsomer} shows the relative ion yields for gas-phase PDO (panel~a) and PDO droplets (panel~b). 
As mentioned above, the high abundance of protons in the droplet ion spectra is a feature that was also observed in reaction nanoscopy studies with (water-covered) silica nanoparticles\,\cite{Rupp2019,Alghabra2021}. We believe that the strong proton signal from droplets is the consequence of the ionization of the droplet by the ultrashort few-cycle laser pulse. Within a very short period of time, the droplet is left in a highly charged non-equilibrium state and the remaining electrons redistribute. A comparison of the electronegativities of the constituents of PDO\,\cite{Allen1989} suggests that proton emission is favorable for the reduction of surface charge and the minimization of the total energy of the droplet.

Another abundant ion in the droplet data is the H$_2^+$ molecular ion. While it is likely also produced from gas-phase PDO to some extent, the signal is below the noise floor, which is why it is not listed in \autoref{Fig:FigureIsomer}a. We assume that the production of H$_2^+$ is greatly enhanced by the proton-rich environment around the nanoparticle surface after the ionization.
As for the protons, the H$_2^+$ molecular ion has been observed in reaction nanoscopy experiments on solid nanoparticles\,\cite{Alghabra2021}. This supports the hypothesis that its production is rather related to the proton-rich environment on the charged nanosurface than the specific material of the nanosphere. 
Our observation of protonated PDO molecules ($m=77$\,u) from droplets is also consistent with this hypothesis.  In fact, for the droplets, the emission of protonated PDO outweighs the emission of PDO$^+$ ions by far.

Finally, we analyze two fragmentation channels of PDO$^+$ that are strongly related to each other as they only differ by the final placement of a proton. These are the loss of a neutral water molecule from PDO$^+$ (resulting in $m=58$\,u) and the emission of an H$_3$O$^+$ ion ($m=19$\,u). Independent of the specific PDO isomer, we find that the ratio between H$_2$O-loss and H$_3$O$^+$ ion production decreases when going from the gas-phase to the droplet. 
This means that the emission of neutral water molecules becomes less favorable on the droplet, while the emission of H$_3$O$^+$ (i.e. protonated water) increases.
This finding is also consistent with a proton-rich environment at the droplet surface which influences the ion generation and the dissociation pathways.

These results demonstrate that many of the ion fragments observed in reaction nanoscopy on droplets appear to be generated similarly to positive-mode electrospray ions, where protonation is known to be the main ion generation mechanism\,\cite{Banerjee2012}.

\subsection{Isomer comparison}
The two isomers under investigation, 1,2-PDO and 1,3-PDO show overall a similar fragmentation behavior for both, the gas-phase and the droplet ion data.
However, the production of one fragment, namely CH$_3^+$, is greatly enhanced for 1,2-PDO droplets as compared to droplets of the other isomer.
For the gas-phase data, the ratio of the relative yields of CH$_3^+$ from 1,2-PDO and 1,3-PDO is about 2:1 (\autoref{Fig:FigureIsomer}a, blue rectangle). The lower production of CH$_3^+$ from 1,3-PDO in this case is likely a consequence of the absence of a methyl group in the 1,3-PDO isomer, which makes the production of CH$_3^+$ a more complex process as compared to 1,2-PDO.
A more pronounced difference between the two isomers is observed in the droplet ion data. 
The CH$_3^+$ yield ratio between the isomers increases to a ratio of 6:1 for droplet ion emission compared to the gas-phase case (\autoref{Fig:FigureIsomer}b, blue rectangle).

To better understand this effect, we carried out density functional theory calculation for both isomers.
\autoref{Fig:FigureIsomer}c shows the minimum energy configuration of the 1,2-PDO interface obtained from DFT. We observe that the methyl groups are pointing away from the interface towards the vacuum side. The three topmost layers of the simulation cell are shown in the inset of \autoref{Fig:FigureIsomer}c where the grey dotted lines represent the hydrogen bonding between neighboring molecules. 
The equilibrium configuration is a result of a steric effect between the methyl groups, which raises the energy, and hydrogen bonding between OH groups, which stabilizes the system. These two effects compete with each other, resulting in the methyl groups pointing outward. See the supplementary material for further details on the steric effect in 1,2-PDO. 
The equilibrium interface for 1,3-PDO is shown in \autoref{Fig:FigureIsomer}d. Unlike its isomer, 1,3-PDO forms a linear chain, linked by hydrogen bonds between the OH groups.

These theoretical results suggest that the experimentally observed difference in the CH$_3^+$ production from both isomers and especially the enhancement for droplets is a direct consequence of the molecular alignment on the droplet surface and the intermolecular hydrogen bonds.

\section{Conclusion}
In this study, we have demonstrated that the reaction nanoscopy technique can be applied to examine the properties and the chemical composition of droplets.
Using droplets of propanediol, we have characterized the average droplet size in our experiment by measuring the near-field tilt using the correlation between the ion emission and the optical near-field on the droplet surface.
We have shown that the measured ion energy can be used to estimate the droplet charge.
We find that in all experiments, the total droplet charges were significantly lower than the Rayleigh limit, indicating stable conditions at the instance of ion emission.
In addition to the physical characterization of the droplets as a whole, we have analyzed the relative yield of ions emitted from propanediol droplets and gas-phase propanediol.
We found evidence for a strong influence of the proton-rich droplet surface on the emitted ion species. 
Moreover, we observed a threefold increase of the CH$_3^+$ production on 1,2-propanediol droplets compared to 1,3-propanediol droplets. 
Density functional theory revealed that this enhancement can be explained by the molecular alignment of 1,2-propanediol molecules on the droplet surface.
These results show that reaction nanoscopy is a versatile tool for studying ion emission from laser-ionized droplets.
With its combination of spatial resolution due to the near-field sensitivity and time resolution due to the use of ultrashort-laser pulses, reaction nanoscopy may enable the spatio-temporal observation of surface reactions on droplets in the future, opening up new possibilities beyond conventional methods like electrospray mass spectrometry.

\paragraph*{Author contributions:}
P.R., R.D. and W.Z. carried out the experiments.
P.R. analyzed the experimental data.
A.M. provided the DFT calculations.
M.N. operated the laser system.
P.R. and A.M. drafted the manuscript.
M.I., B.B., and M.F.K. supervised and coordinated the project. 
All authors reviewed the manuscript and contributed to its completion. 

\begin{acknowledgement}
The authors thank A.J. Feinberg for proofreading the manuscript and M. Gra{\ss}l for his feedback on the figures.
\end{acknowledgement}

\begin{funding}
The experiments were carried out at LMU, where the
work was supported by the German Research Foundation (DFG).
M.F.K. is grateful for partial support by the Max Planck Society via the Max Planck Fellow
program.
M.F.K.’s work at SLAC is supported by the U.S. Department of Energy, Office of Science, Basic Energy Sciences, Scientific User Facilities Division and by the Chemical Sciences, Geosciences, and Biosciences division with award DE-SC0063. 
W.Z. acknowledges support from the Alexander von Humboldt Foundation and the MULTIPLY fellowship program under the Marie Skłodowska-Curie COFUND Action.
A.M. and M.I. acknowledge the support by the U.S. Department of Energy, Basic Energy Sciences, Gas Phase Chemical Physics program with award DE-SC0021129.
\end{funding}

\printbibliography{}

\end{document}


\maketitle

\section{Simulation geometry}
\begin{figure}[hb]
    \centering
    \includegraphics[width=0.6\columnwidth]{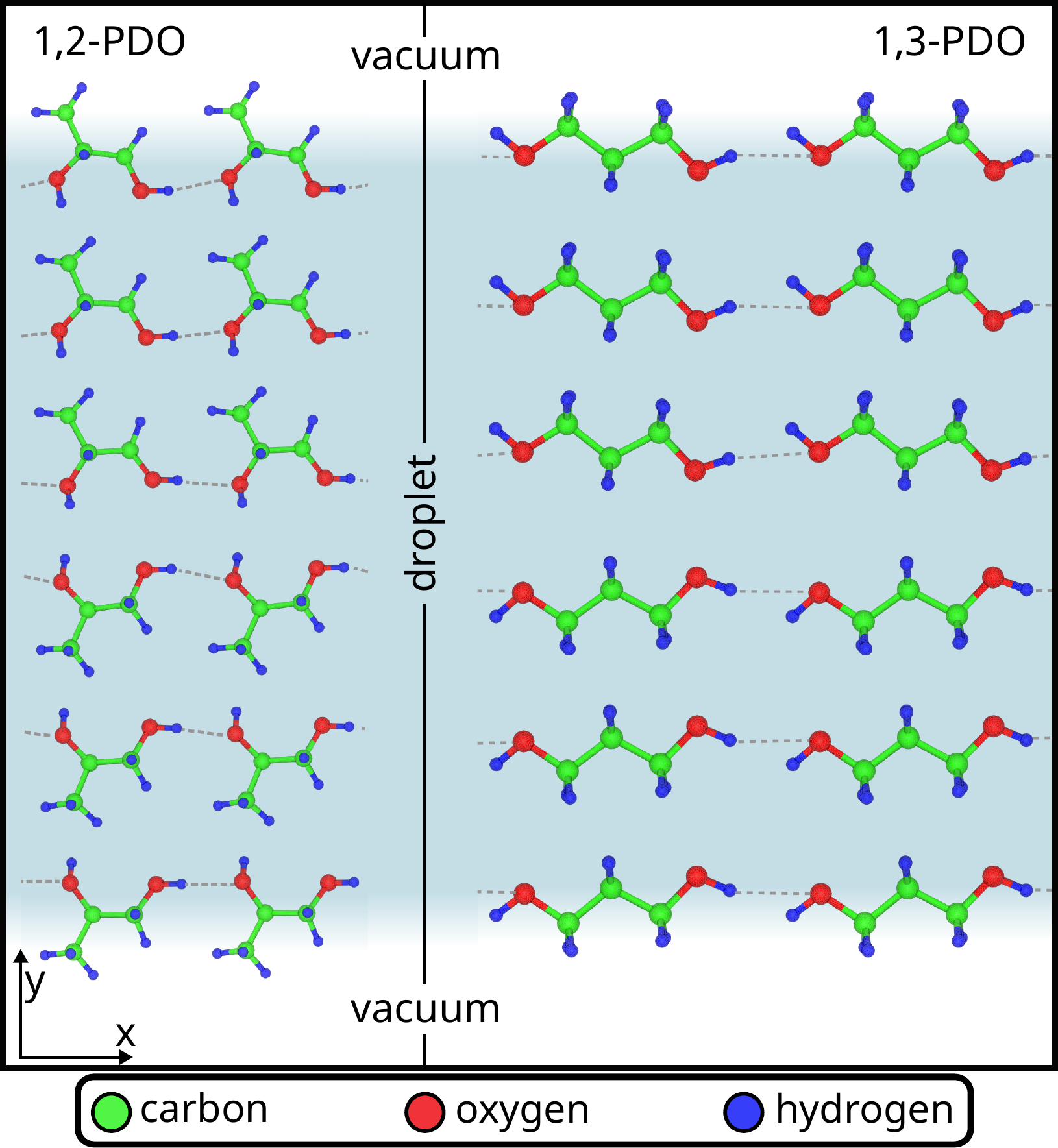}
    \caption{Side view of the simulation geometry for 1,2-PDO (left) 1,3-PDO (right). 
    The surface normal is along the $y$-axis. Periodic boundary conditions were applied along $x$ and $z$. The dotted lines represent hydrogen bonds.}
    \label{fig:simgeom}
\end{figure}

The interface between the nanodroplet and vacuum has been modeled using a slab of molecules in the $x$-$z$ plane consisting of six layers as shown in \autoref{fig:simgeom}. For an unambiguous definition of the droplet-vacuum interface, the bottom three layers are a rotated image of the top layers. The formation enthalpy for the 1,2-PDO and 1,3-PDO slab from isolated gaseous molecule is -16.14\,kcal/mol (0.7\,eV) and -31.14\,kcal/mol (1.35\,eV), respectively. This suggests that simulation configurations are energetically stable.

\newpage

\section{Ion fragment selection}
\begin{figure*}[h]
	\centering\includegraphics[width=\textwidth]{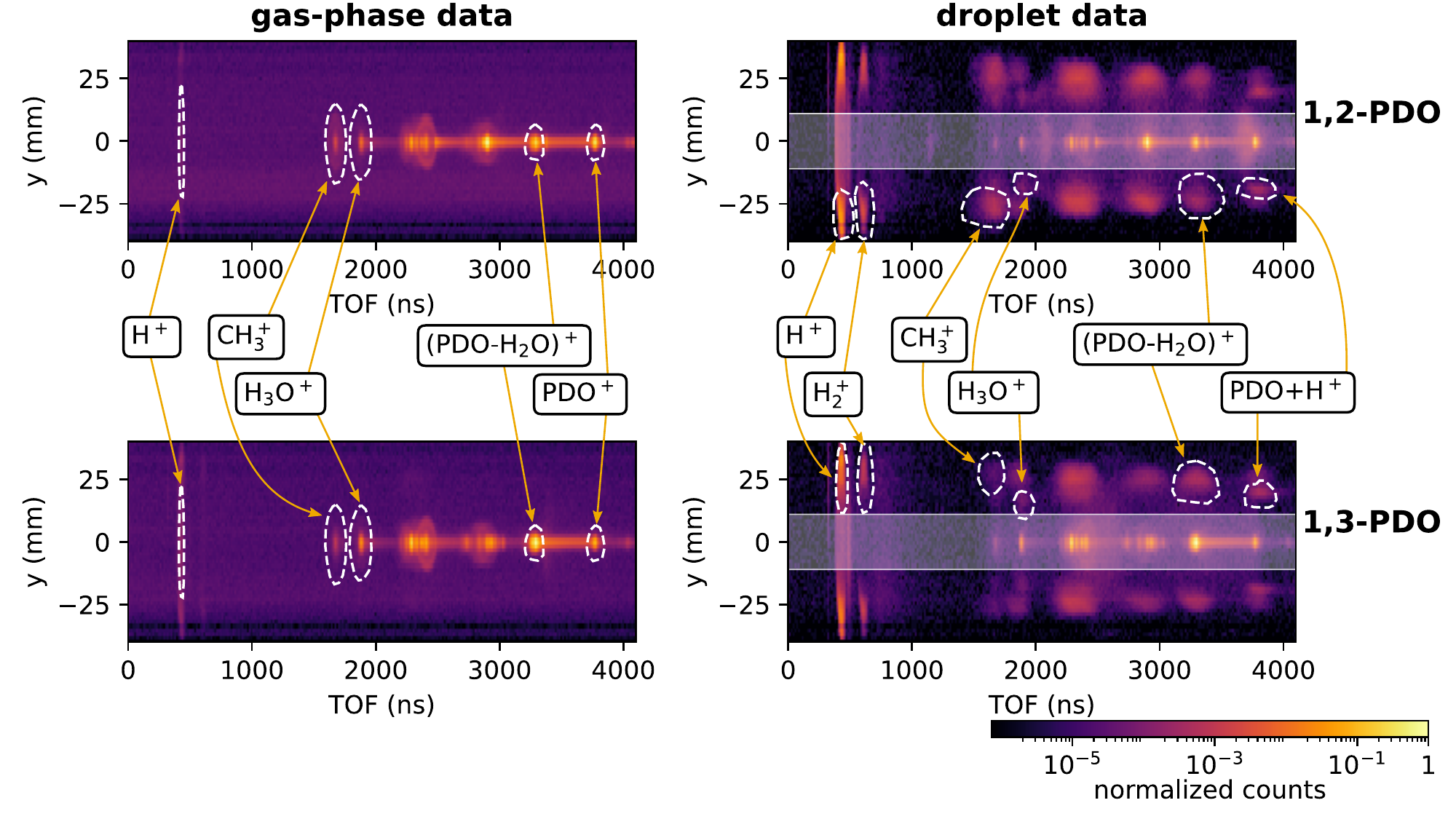}
	\caption{\label{Fig:FragmentSelectionSI} Filtering ion signals for gas-phase and droplet data. 
	Two measurements are shown, one for 1,2-propanediol (1,2-PDO, panels a and b) and one for 1,3-propanediol (1,3-PDO, panels c and d). 
	Both measurements were split into gas-phase data and droplet based on the strength of the electron signal which is recorded in coincidence with the ions. 
	Gas-phase data (first column, panels {\bf a} and {\bf c}) are defined by a small or entirely absent electron signal. Droplet data (second column, panels {\bf b} and {\bf d}) are defined as the complementary set. 
	The ion species listed in the middle were identified after calibrating the time-of-flight. The polygons indicated by the white dashed lines were used for the yield comparison in the main text (Figure~5). The white shading corresponds to $|y|<11$\,mm.}
\end{figure*}

\begin{figure}
    \centering
    \includegraphics[width=0.8\textwidth]{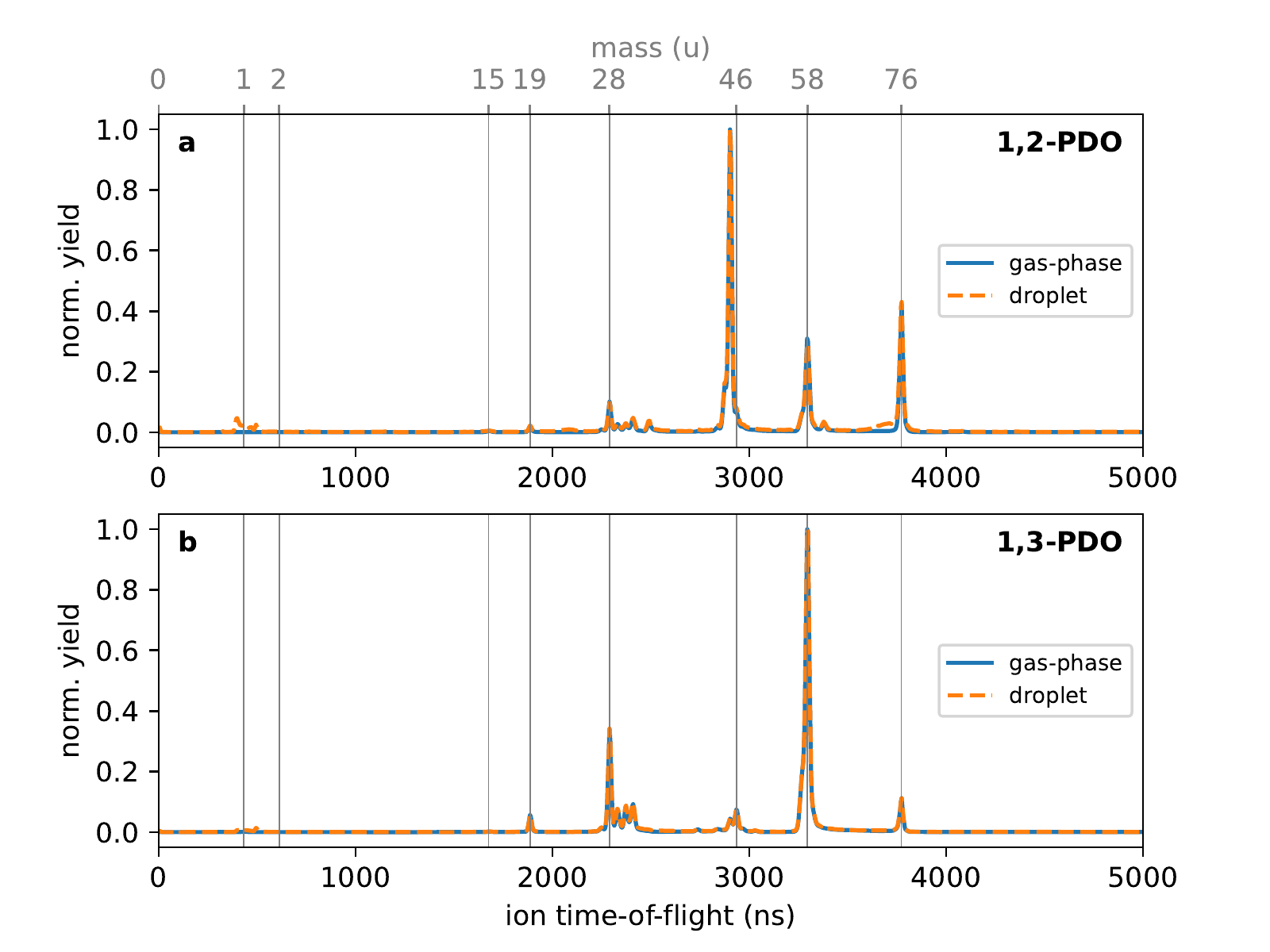}
    \caption{Time-of-flight histograms for $|y|<11$\,mm. The histograms were obtained by projecting the corresponding parts of the position-resolved spectra of \autoref{Fig:FragmentSelectionSI} (for the droplet data, we have indicated $|y|<11$\,mm with the white shading). }
    \label{fig:1D_tof_SI}
\end{figure}

The measurement of the electron signal in reaction nanoscopy enables a significant reduction of the single molecule ionization signal in the droplet ion data. As shown in Figure 1 of the main text, droplet events cause a much larger electron signal than gas-phase events, which allows filtering the data.
Despite the filtering, we always observe some contribution of gas-phase, single molecule ionization to the droplet data.
From a comparison of the gas-phase data and the droplet data presented in \autoref{Fig:FragmentSelectionSI}, it becomes clear that the peaks close to $y=0$ in the droplet data (which show up identically in the gas-phase data) are ions emitted from single gas-phase molecules which were ionized independently from the droplet but by the same laser pulse. For this case of simultaneous ionization of a droplet and single-molecules, the electron signal is still large and the whole laser shot is assigned to the droplet data.
This effect is further confirmed by the 1D time-of-flight histograms in \autoref{fig:1D_tof_SI}, where data for $|y|<11$\,mm is shown, corresponding to the white-shaded regions of 
Figure~2 in the main text and \autoref{Fig:FragmentSelectionSI}.

Similar but fragment-specific filters were applied for the ion data shown in Figure~5 of the main text.
For every ion species under investigation, we defined a polygon in the space of time-of-flight and y-position. These polygons are shown in \autoref{Fig:FragmentSelectionSI}. For the droplet data, all the polygons were additionally mirrored around $y=0$ in order to include both lobes of the dipolar ion emission pattern.

\clearpage
\newpage
\section{Steric effects on the surface of 1,2-PDO droplets}
\begin{figure}[htbp!]
	\centering\includegraphics[width=0.8\textwidth]{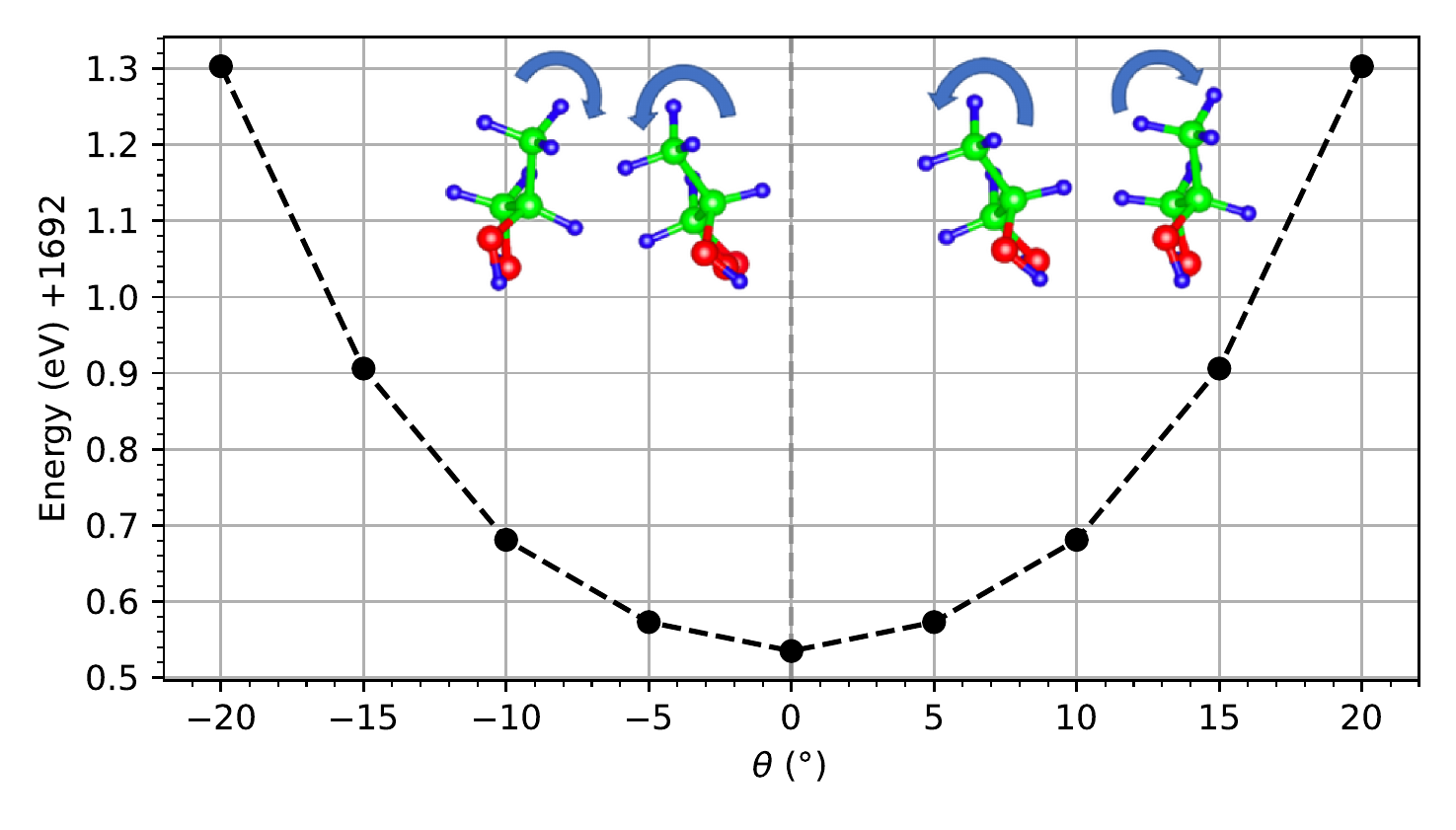}
	\caption{\label{Fig:FigureSteric} Energy of 1,2-propanediol with respect to angle of rotation $\theta$. The rotation axis is along the C2-C3 bond and $\theta=0^\circ$ corresponds to the equilibrium configuration obtained via energy minimization. The rotations are depicted by the blue arrows.}
\end{figure}

To further confirm our understanding about the influence of steric effects between methyl groups on the minimum energy configuration, we performed single point energy calculations of 1,2-PDO by changing the configuration of the molecules from equilibrium. We rotate the molecules located at the interface inward and outward from the minimum energy state by an angle $\theta$ around the C2-C3 bond (\autoref{Fig:FigureSteric}, shown by blue arrows), resulting in the methyl groups of neighboring molecules coming close to each other and thereby raising the energy of the system. This corroborates the fact that the steric effect plays an important role in the surface structure of 1,2-PDO droplets.

